# FEASIBILITY OF OTR IMAGING FOR LASER-DRIVEN PLASMA ACCELERATOR ELECTRON-BEAM DIAGNOSTICS*

A. H. Lumpkin[1#], D.W. Rule[2], and M.C. Downer[3]**
[1]Fermi National Accelerator Laboratory, Batavia, IL 60510 USA,
[2]Silver Spring, Maryland 20904, [3]University of Texas, Austin, Texas 78712

*Abstract*

We report the initial considerations of using linearly polarized optical transition radiation (OTR) to characterize the electron beams of laser plasma accelerators (LPAs) such as at the Univ. of Texas at Austin. The two LPAs operate at 100 MeV and 2-GeV, and they currently have estimated normalized emittances at ~ 1-mm mrad regime with beam divergences less than 1/γ and beam sizes to be determined at the micron level. Analytical modeling results indicate the feasibility of using these OTR techniques for the LPA applications.

## INTRODUCTION

Recent measurements of betatron x-ray emission from quasi-monoenergetic electrons accelerating to 500 MeV within a laser plasma accelerator (LPA) enabled estimates of normalized transverse emittance well below 1 mm-mrad and divergences of order 1/γ, where γ is the Lorentz factor [1]. Such unprecedented LPA beam parameters can, in principle, be addressed by utilizing the properties of optical transition radiation (OTR). In particular, the linearly polarized features of that radiation provide additional beam parameter sensitivity. We propose a set of complementary measurements of beam size and divergence with near-field and far-field OTR imaging, respectively, on LPA electron beams ranging in energy from 100 MeV [2] to 2 GeV [3]. The feasibility is supported by analytical modeling for beam size sensitivity and divergence sensitivity. In the latter case, the calculations indicate that the parallel polarization component of the far-field OTR pattern is sensitive to rms divergences ($\sigma_\theta$) from 0.1 to 0.4 mrad at 2 GeV, and it is similarly sensitive to rms divergences from 1 to 5 mrad at 100 MeV.

We anticipate the signal levels from charges of 100 pC will require a 16-bit cooled CCD or scientific CMOS camera. Other practical challenges of utilizing these techniques with the LPA configurations will also be discussed. These include the fundamental requirement to deflect the high power laser component with a foil while scattering the electron beam less than its intrinsic divergence. This may be achieved with a replaceable foil.

*Work partly supported under Contract No. DE-AC02-07CH11359 with the United States Department of Energy.
**Work at the Univ. of Texas supported by DoE grant DE-SC0011617.
#lumpkin@fnal.gov

## TECHNICAL CONSIDERATIONS

Two main aspects of the proposed experiments are to install OTR stations with near-field and far-field imaging options in the LPAs to assess electron beam size and divergence. A brief summary of the two LPAs at the University of Texas at Austin will be described with current diagnostics, and then the proposed OTR techniques will be addressed.

### *The Laser Plasma Accelerators*

The Texas PW LPA schematic is shown in Fig. 1. The PW laser is focused onto the gas jet of 7-cm extent. At plasma electron densities of 3 to 5 x $10^{17}$ $cm^{-3}$ strong plasma wake field acceleration occurred, and the electrons attained quasi-monoenergetic energies of 2 GeV [3]. Normally a dipole magnet is used to provide a dispersive effect in the x-plane for energy and energy spread measurements as detected by a downstream LANEX phosphor screen, imaging plate, or other. The vertical beam divergence is measured by evaluating the vertical beam size at the known drift location. A summary of the beam parameters for the PW LPA is given in Table 1. A similar process occurs with a TW LPA driven by a 30 TW laser to provide energies of 100 MeV [2], or 0.1 GeV, as summarized in Table 2.

Table 1: Summary of some PW LPA electron beam properties.The estimated parameters are indicated*.

| Parameter | Units | Value | Range |
|---|---|---|---|
| Energy | GeV | 2 | 1-2 |
| Emittance* | mm mrad, norm. | 1 | 0.1-2 |
| Beam size* | μm | 1 | 0.1-2 |
| Divergence | mrad (FWHM) | 0.3 | 0.2-0.4 |
| Charge | pC | 100 | 10-100 |
| Duration* | fs | 10 | 10-30 |

Table 2: Summary of some TW LPA electron beam properties. The estimated parameters are indicated*.

| Parameter | Units | Value | Range |
|---|---|---|---|
| Energy | GeV | 0.100 | .050-0.100 |
| Emittance* | mm mrad, norm. | 1 | 0.1-1 |
| Beam size* | μm | 1 | 0.1-2 |
| Divergence | mrad FWHM | 4 | 2-5 |
| Charge | pC | 100 | 10-100 |
| Duration* | fs | 10 | 10-30 |

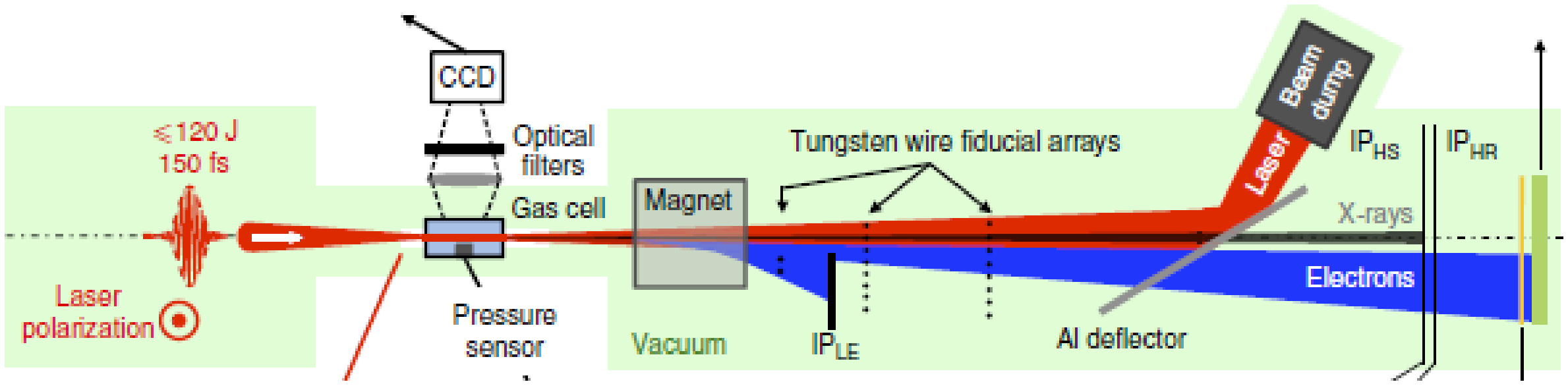

Figure 1: Schematic of the UT PW laser plasma accelerator showing the laser pulse, gas cell, dipole magnet, Al deflector for laser, and imaging screens/plates. This LPA generated 2-GeV electron beams [2]. Measurement stations for OTR would be just outside of the plasma bubble regime as well as downstream.

### *OTR Basics*

Optical transition radiation is emitted at the boundary of two media when relativistic charged particles induce currents in the media [4-7]. Both forward and backward OTR are generated as schematically shown in Fig. 2a. In addition the backward OTR is emitted in a cone around the angle of specular reflection as illustrated in Fig. 2b, and the dipole lobes are folded into the $1/\gamma$ opening angles. For high gamma, these angles are much smaller than the opening angle of Cherenkov radiation as in the example with an index of refraction, n. These basic OTR features will be utilized in the LPA application for setting the foil angles and the optics.

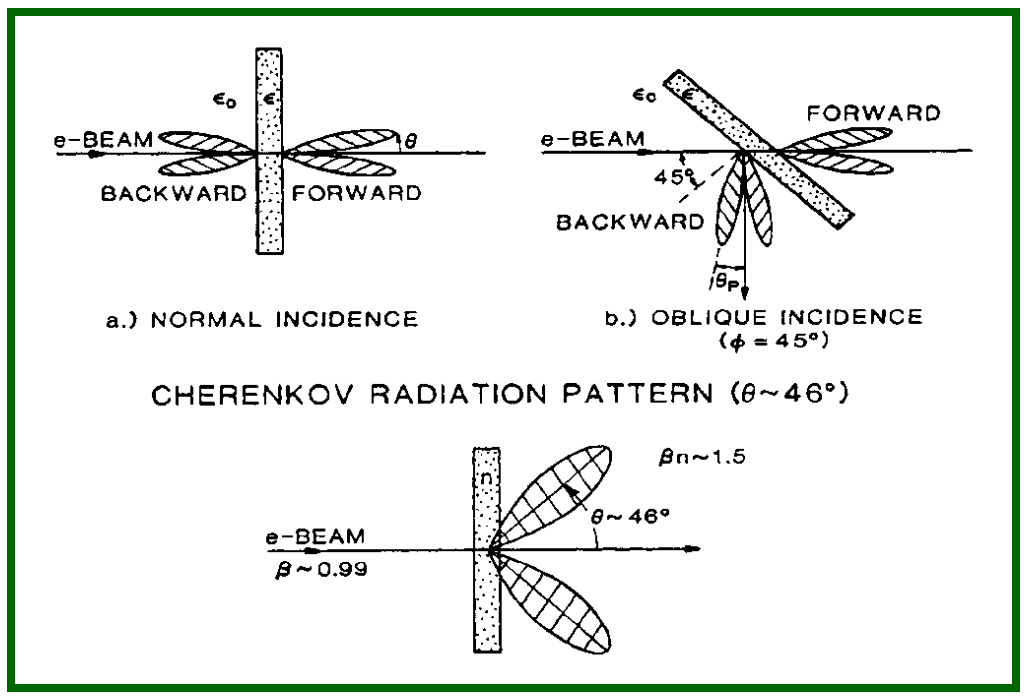

Figure 2: OTR schematic showing a) forward and backward lobes, b) the oblique-angle case, and c) a Cherenkov case with theta=46 degrees for βn=1.5.

The ability to measure both the beam size and the divergence as separable parameters is realized by imaging OTR in the near field and the far field, respectively, as illustrated in Fig. 3. The beam-size measurement is made with the sensor at the image plane, and the angular distribution measurement is made with the sensor at the focal plane of the lens system. The angular distribution pattern carries energy information in the opening angle, divergence in the visibility of the central minimum, and pointing angle in the centroid of the pattern.

The modeling of the OTR point spread function (PSF) [8-12] used in beam-size imaging and the OTR angular distribution [6,7,13] have been described previously. We will report the results of our modeling in the next section.

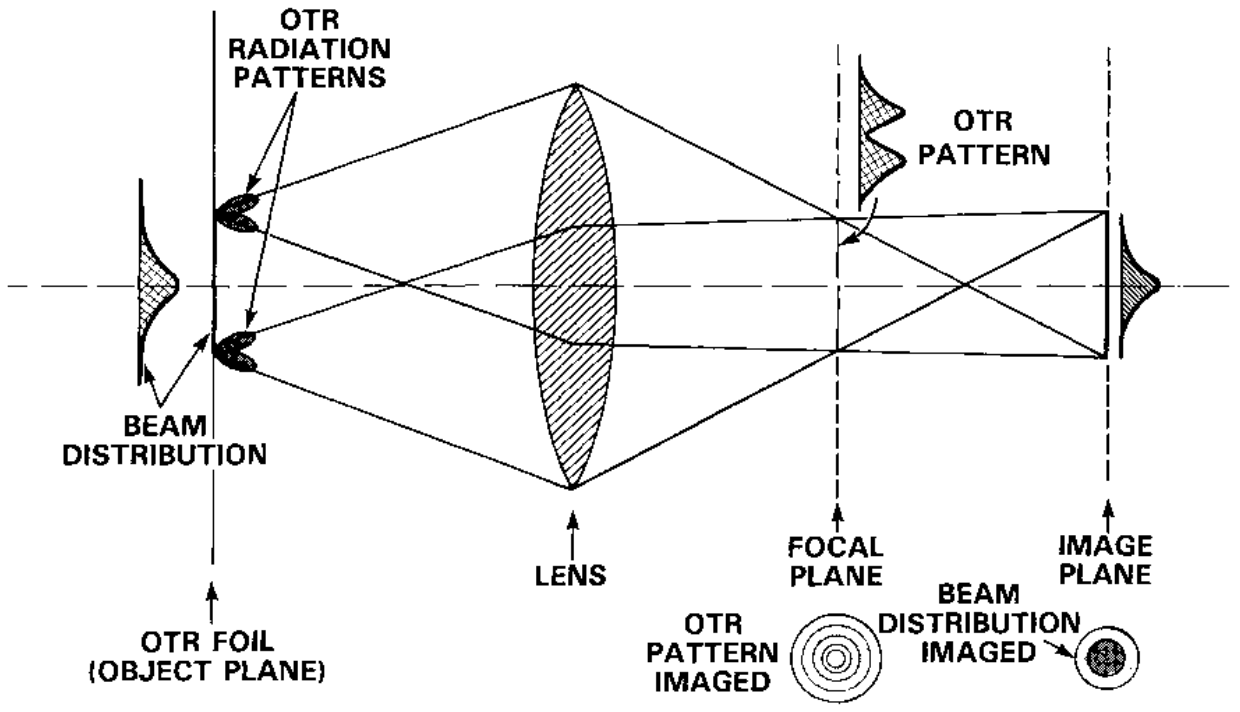

Figure 3: Schematic of near-field and far-field OTR imaging which provide the beam distribution and angular distribution, respectively.

## MODELING RESULTS

### *Near-field Imaging for Beam Distributions*

We show the results of the OTR–PSF model for a maximum collection angle of 250 mrad, optical magnification of 10, and at 500 nm wavelength in Fig. 4. The total, horizontal polarization, and vertical polarization images are indicated. Note, these are in the image plane. One can visualize that the horizontal polarization with horizontal projection is double lobed while the corresponding vertical projection would be single lobed (and thus provide better effective resolution). The projections are shown in Fig. 5a. This particular case is for 2 GeV and for a beam size of 1 µm. In Fig. 5b we show the result of convolving the respective PSF polarization components with the beam size, and the smallest observed size is 1.6 µm using the horizontal polarization with vertical projection, effectively. We have performed similar calculations with smaller beam sizes and for the 100-MeV cases which display an additional feature that was beam-size sensitive. This is a subject for further investigations. In practice, one would deconvolve the PSF from the observed profile to obtain the original.

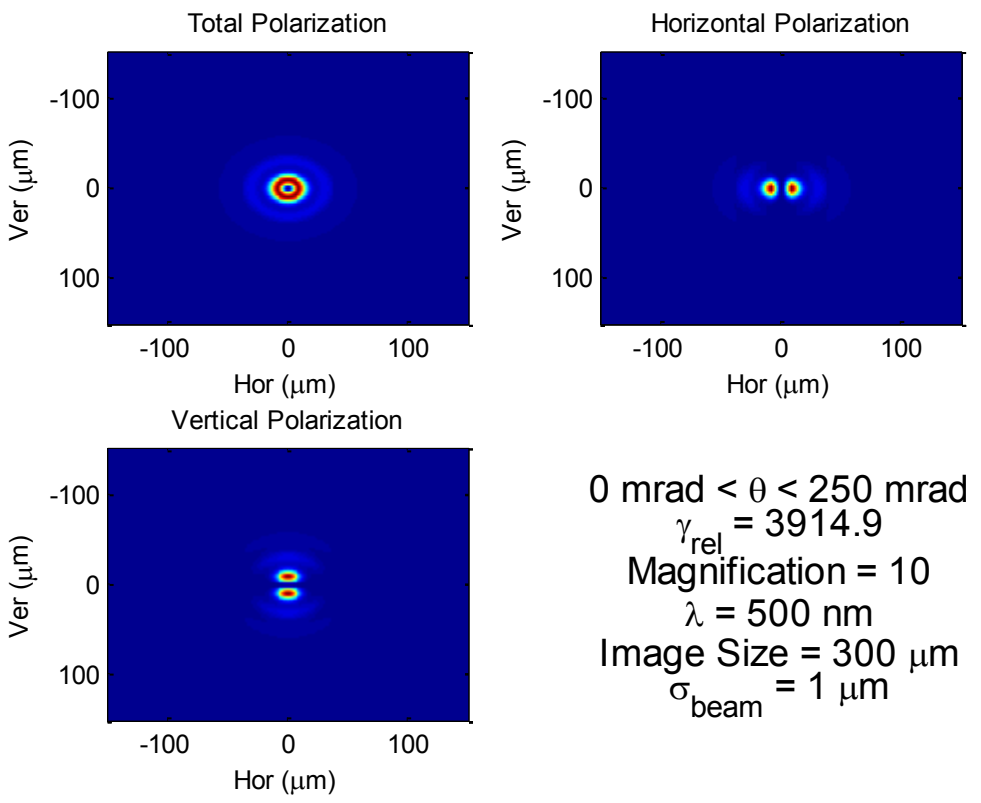

Figure 4: Calculated PSF distributions in the image plane with theta max, magnification, and wavelength indicated.

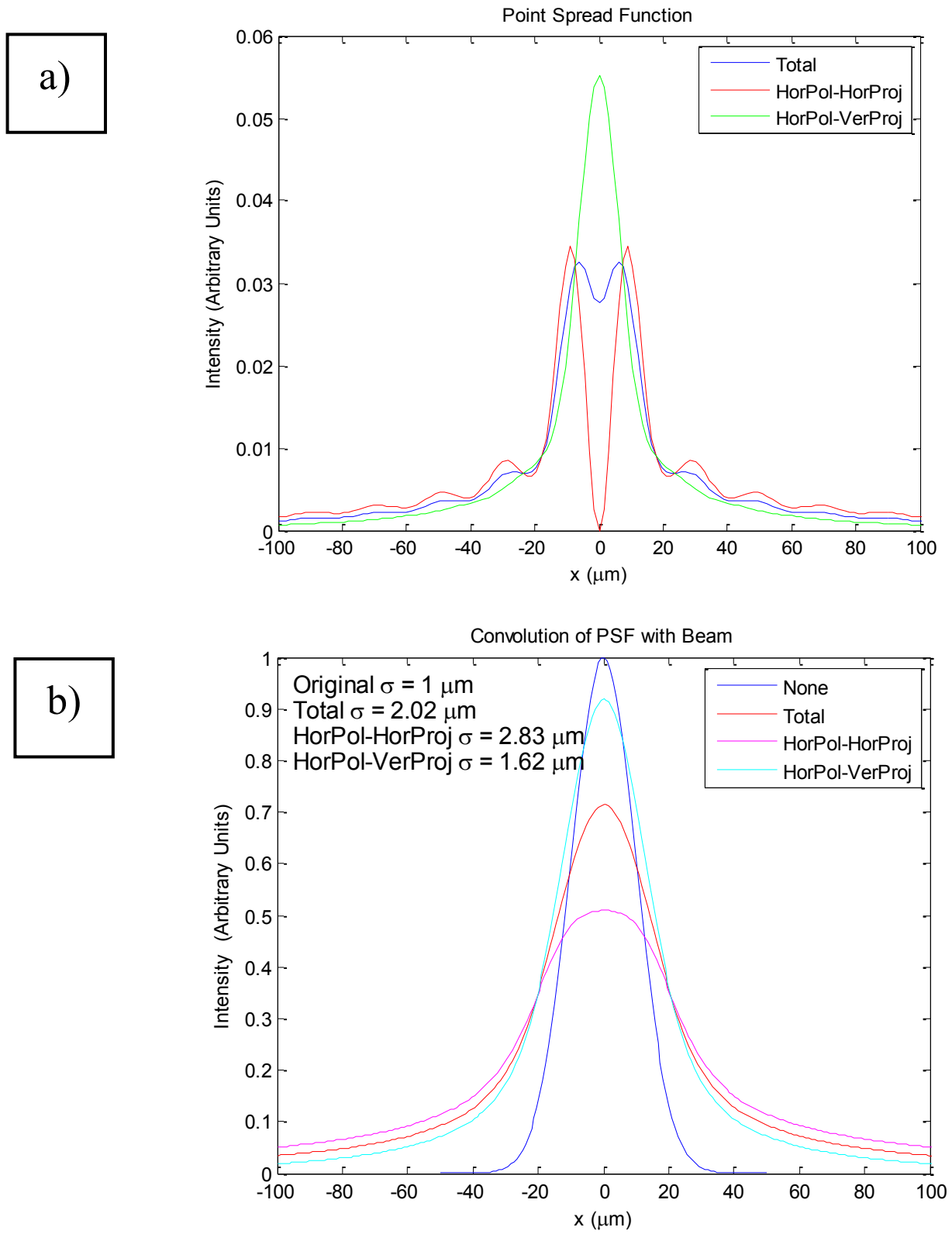

Figure 5: Projections of a) the PSF polarized components and b) the convolutions with the 1-µm beam size.

## *Far-field Imaging for Angular Distributions*

We consider the angular distribution pattern calculated for the 2-GeV case with 0.1-mrad rms divergence in Fig. 6 and 0.2-mrad rms divergence in Fig. 7. We show three intensity/polarization components: perpendicular (Iperp), parallel (Ipar) and the total (Itot). The reference plane is formed by the e-beam direction and the observation angle. The reduction of the Ipar modulation between the lobes with increased divergence is clear. Under a Gaussian distribution assumption, we have divided the reported FWHM value in Table 1 by 2.35 so our cases are close. In Fig. 8 we show a result for the 100-MeV beam with 2-mrad divergence. In the 4-mrad divergence case, Ipar modulation was markedly reduced (not shown here).

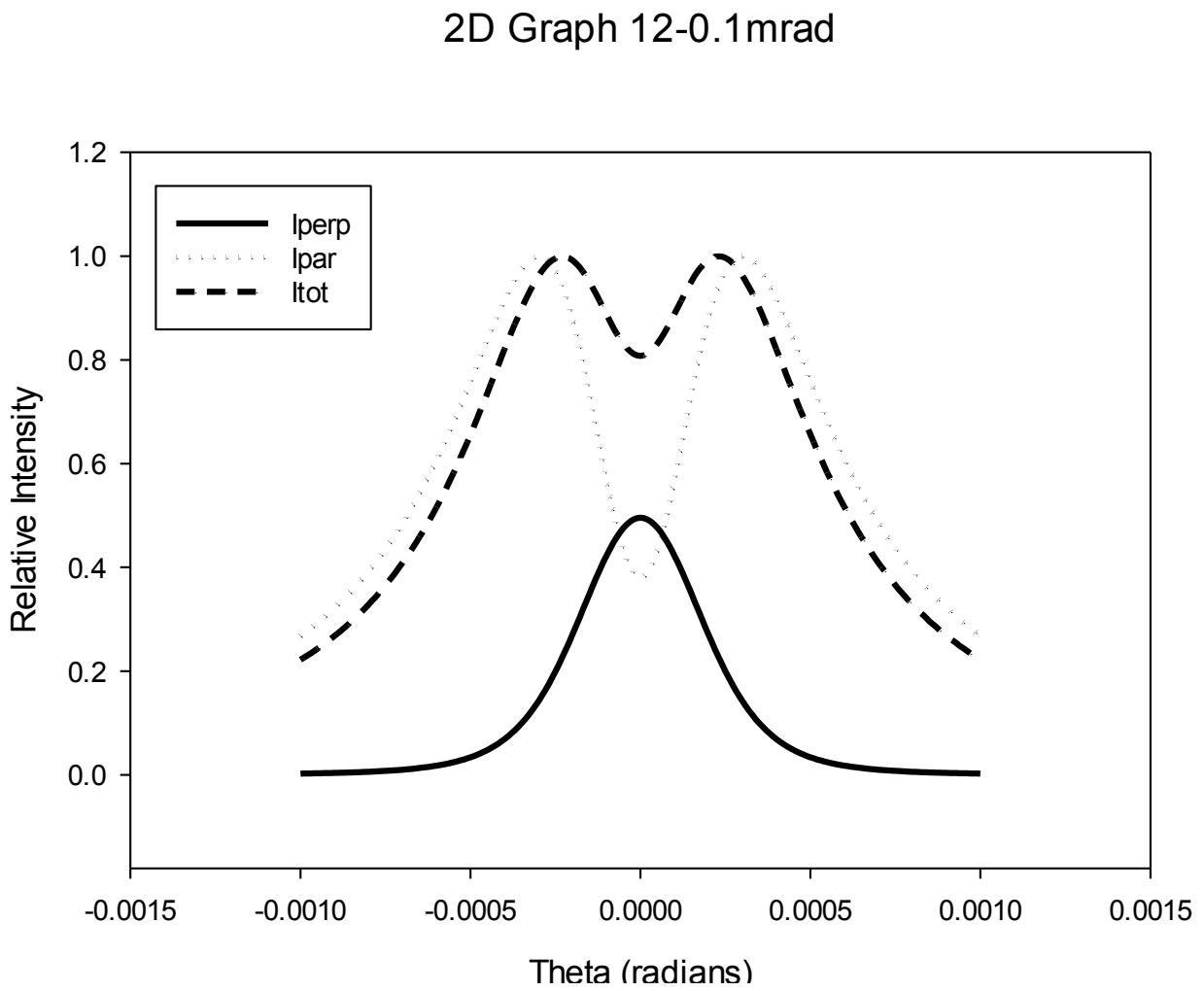

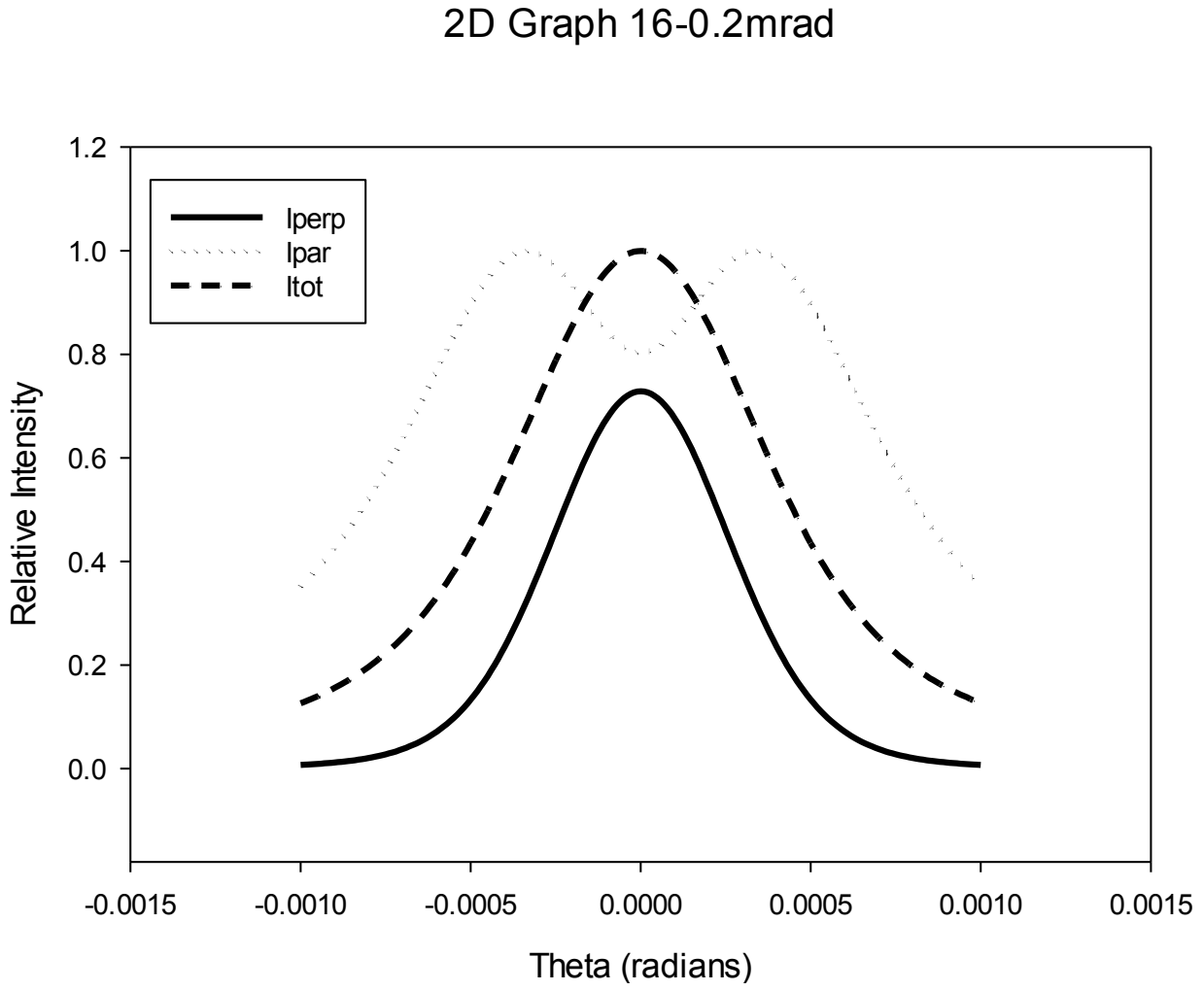

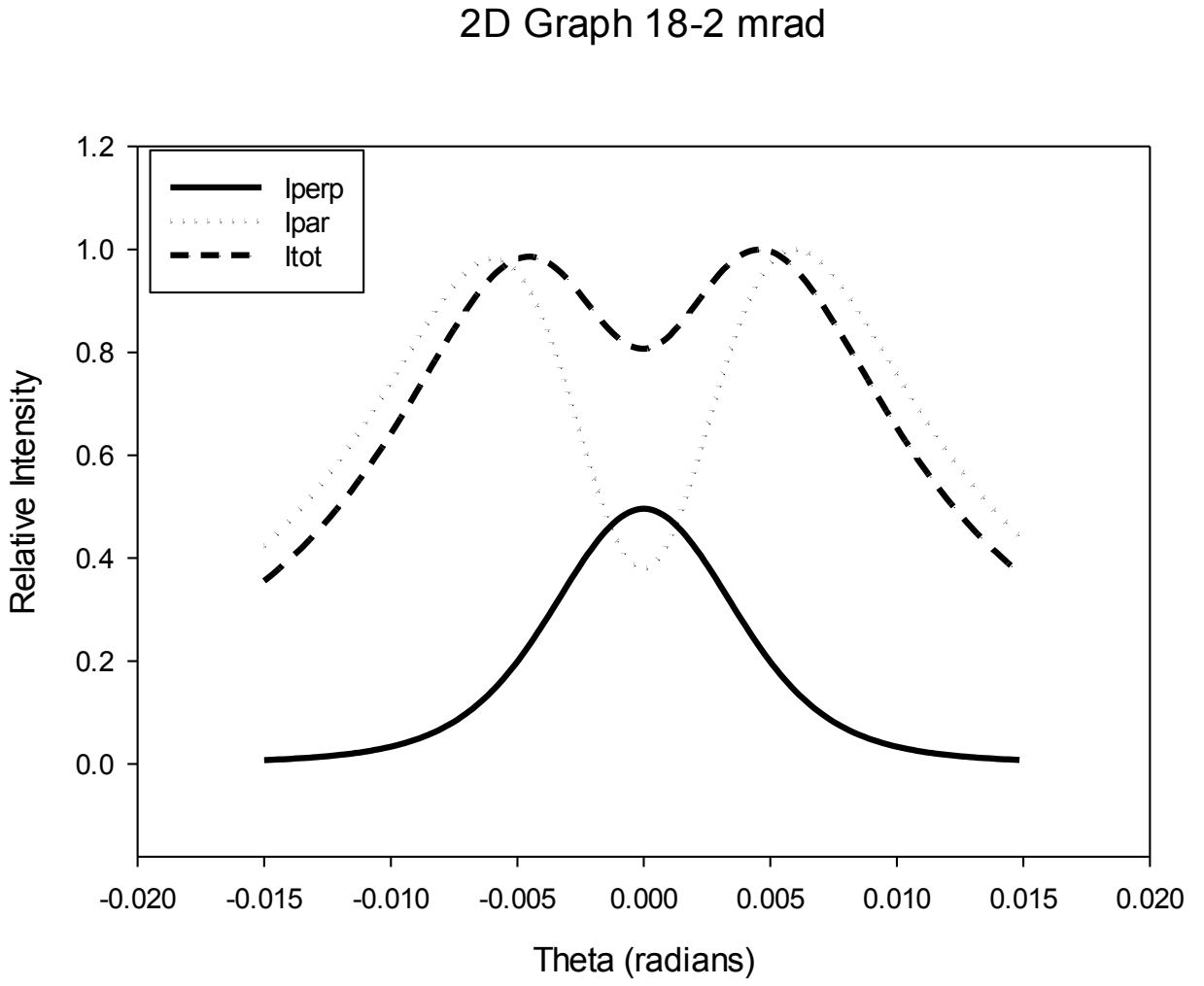

Figure 8: Calculated intensity projections of the far field image for 0.1-GeV beam energy and 2-mrad divergence.

# SUMMARY

In summary, we have done preliminary evaluations of the sensitivity of polarized OTR imaging to beam size and divergence for the two LPAs at UT-Austin. For the values considered, the feasibility was established for both parameters and both LPAs. Experiments are being planned in the next phase to implement these techniques.

## ACKNOWLEDGMENTS

The first author acknowledges the support of R. Dixon and N. Eddy at Fermilab.

This research is dedicated *in memoriam* to Helen Edwards.